# Engineering Knowledge Graph from Patent Database

L Siddharth[12*], Lucienne T.M. Blessing[12], Kristin L. Wood[3], Jianxi Luo[12]

[1]*Engineering Product Development Pillar, Singapore University of Technology and Design, 8 Somapah Road, Singapore, 487372*

[2]*SUTD-MIT International Design Centre, Singapore University of Technology and Design, 8 Somapah Road, Singapore, 487372*

[3]*College of Engineering, Design, and Computing, University of Colorado Denver, 1201 Larimer St, Denver, CO, USA, 80204*

## Abstract

We propose a large, scalable engineering knowledge graph, comprising sets of <entity, relationship, entity> triples that are real-world engineering 'facts' found in the patent database. We apply a set of rules based on the syntactic and lexical properties of claims in a patent document to extract facts. We aggregate these facts within each patent document and integrate the aggregated sets of facts across the patent database to obtain the engineering knowledge graph. Such a knowledge graph is expected to support inference, reasoning, and recalling in various engineering tasks. The knowledge graph has a greater size and coverage in comparison with the previously used knowledge graphs and semantic networks in the engineering literature.

[*] Corresponding author.
Email: siddharth_l@mymail.sutd.edu.sg

# 1. Introduction

Knowledge graphs often serve as the knowledge base and anchor the reasoning facet of Artificial Intelligence – AI [1]–[3], which governs several applications such as search, recommendation, Q&A etc. These applications are continuously being built using the common-sense knowledge graphs that are both open-source (e.g., ConceptNet[1]) as well as proprietary (e.g., Google[2], Microsoft[3]). The vibrant research and development that concerns knowledge graphs indicates a huge potential for knowledge-based AI in engineering.

In a knowledge graph, a real-world fact could be expressed as a triple – <h, r, t>, e.g., <'Anfield', 'is a', 'Football Stadium'> and a set of such facts together constitute a graph network, which supports reasoning tasks [4]. For instance, as shown in Figure 1A, if Anfield is a football stadium and it is located in Liverpool, we could infer that Liverpool has a football stadium. Similar inferences could be drawn from engineering facts as well. As shown in Figure 1B, a resistor heats glue, which transforms into molten glue that when pushed by the actuator, ejects from the nozzle. From this example, we could infer that the resistor produces molten glue, and the actuator is connected to the nozzle.

---

[1] https://conceptnet.io/
[2] https://developers.google.com/knowledge-graph
[3] https://www.microsoft.com/en-us/research/project/microsoft-academic-graph/

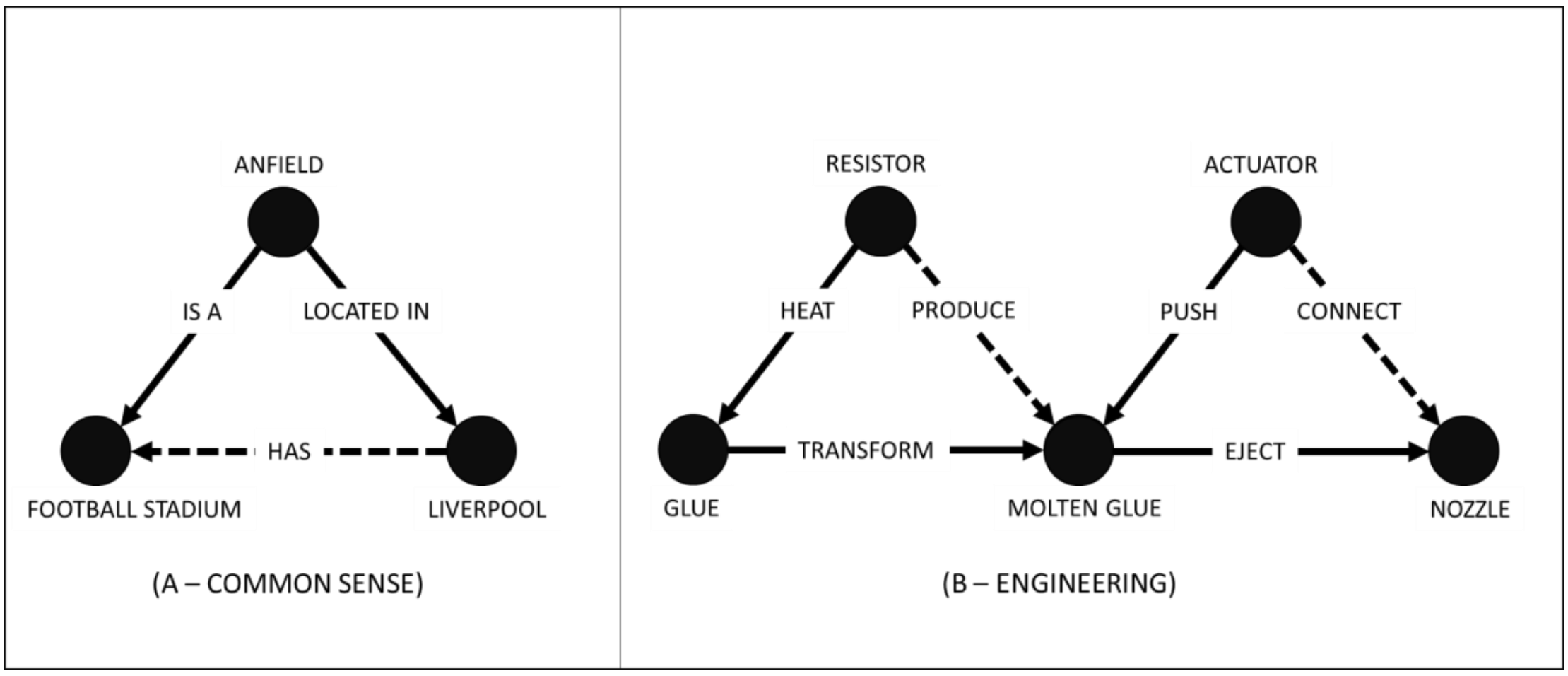

Figure 1: Illustrating inferences in a knowledge graph.

The inferred relationship between nozzle and actuator must be considered when design changes are made to either of these components [5]. Such inferences made over facts in a knowledge graph are also useful for root-cause analysis [6] and synthesis [7]. Despite the advantages of a knowledge graph representation, the design knowledge, however, is often represented using networks [8] and models of causality [7] that deliver abstractions rather than facts.

To meet the growing demands for engineering knowledge retrieval, representation, and concept generation-cum-evaluation, scholars have resorted to large scale, cross-domain, and common-sense knowledge graphs like ConceptNet and FreeBase [9]–[13] that may not provide facts that are technically dominant or relevant. Meanwhile, there is a lack of engineering-contextualized knowledge graph at the scale of ConceptNet and alike [14]. Our work aims to depart from the reliance on common-sense knowledge graphs by developing an engineering knowledge graph.

To create a knowledge graph for engineering, we sought the US Patent database, which is currently the most structured, accessible engineering design database that is developed using natural language text with consistent document structure [15], [16]. We leverage the syntactic and lexical properties of the patent claims [17], [18] to develop and apply some rules to automatically extract facts from these, which are combined to create knowledge graph representations of individual patents. These representations, when integrated, form a large knowledge graph database for engineering.

The core contribution of our work is a large scalable engineering knowledge graph that comprises millions of heterogenous <entity, relationship, entity> triples from the engineering design text data. It is expected to bridge the gap between knowledge and design engineering and enable many research and development opportunities on knowledge reasoning, retrieval, and representation in the context of engineering.

## 2. Background

The research into knowledge graphs generally involves creation, completion, and classification. Creation involves Natural Language Processing (NLP) techniques to extract entities and relationships from natural language text [19]. Completion uses embedding methods like TransE to predict the missing links in a knowledge graph [20]. Classification and/or characterization of nodes (entities) uses the embeddings of these as inputs to classifiers like Graph Neural Networks – GNNs [21]. Our research only focuses on the creation of knowledge graphs, i.e., to extract these from natural language text.

DBpedia[4] is a knowledge graph that was populated using crowdsourced effort to extract nearly 400 million facts from 3.7 million entities found in Wikipedia text [22]. Open Mind Common Sense is a crowdsourcing project that aimed to populate facts for ConceptNet [23], which includes 21 relationships like ‘is a’, ‘part of’ etc. ConceptNet borrows lexical relationships from Wiktionary[5] and ontology from Cyc[6], which is also a knowledge graph.

[4] https://www.dbpedia.org/

Currently, the largest knowledge graph is that of Google – including over 70 billion facts that aids queries using Google Search, Google Assistant, Google Home, and Google Developer API[7]. Some entity types in the Google Knowledge Graph include MusicAlbum, LocalBusiness, MovieSeries, EducationalOrganization etc. Similar to Google’s knowledge graph, Amazon Alexa is aided by Evi[8].

In the engineering literature, scholars have often relied on WordNet – a lexical database[9], which provides both lexical (e.g., hypernym) and quantitative (e.g., Jiang-Conrath similarity [24]) relationships between common-sense terms [11], [25]–[31]. Han et al. [9], [32] utilise ConceptNet relationships to obtain analogies and combinations for a search entity. To evaluate crowdsourced design ideas and extract entities from these, Camburn et al. [12] use the TextRazor[10] platform that is built using models trained on DBPedia, Freebase etc. Chen and Krishnamurthy [10] facilitate human-AI collaboration in completing problem formulation mind maps with the help of ConceptNet and the underlying relationships. These common-sense knowledge bases utilised by the scholars, however, were not built for the engineering purposes.

To develop knowledge graphs for engineering, scholars have adopted a variety of approaches. Yamamoto et al. [33] extract <noun, part of, noun> triples (e.g., “wheel of car”) using the ESPRESSO algorithm [34] and replace the nouns with the adjacent verbs to obtain a functional decomposition model. Park et al. [35] quantify triples like “fabric + shrink” using Wu and

[5] https://en.wiktionary.org/wiki/Wiktionary:Main_Page
[6] https://www.cyc.com/archives/service/cyc-knowledge-base
[7] https://developers.google.com/knowledge-graph
[8] https://www.evi.com/
[9] https://wordnet.princeton.edu/
[10] https://www.textrazor.com/

Palmer similarity [36] connect these using explicit causal conjunctions like ‘so’, ‘due to’, ‘because’ etc [35].

Li et al. [37] extract a large knowledge graph by mining SVO tripes of the form: $\underbrace{NP_{sub}}_{amod} \xrightarrow{nsubj} \underbrace{VP}_{advmod} \xrightarrow{dobj} \underbrace{NP_{obj}}_{amod}$ from 18,000 Chinese websites that relate to healthcare. Using experts, they also construct a nursing bed knowledge graph based on the FBS model. Using these internal and external knowledge graphs, based on C-K theory [38], they demonstrate how concepts are generated using knowledge graphs, which expands as new concepts are generated. Upon extracting Subject-Verb-Object (SVO) triples from Wikipedia articles, Cheong et al. [39] classify verbs and objects into functions and flows on a functional basis using Jiang-Conrath similarity [24] and Word2Vec [40] similarities against the term – ‘energy’.

None of the approaches reviewed so far provide evidence of building knowledge graphs out of natural language text that is purely documented for engineering. Hence, the existing knowledge graphs built by the engineering scholars are less likely to deliver pure technical knowledge. Our research fills this gap by extracting knowledge graphs from claims in patent documents that were granted by the United States Patent and Trademark Office (USPTO). Although scholars have mined the patent documents [16], [41]–[46], claims in these remain unexplored in data-driven engineering design research. We chose claims as our data source due to their syntactic-cum-lexical properties [18] and the existence of triples [17] in these. The properties of claims that are relevant to our work are listed as follows.

1. Each claim is written in a single sentence.
2. The dependencies between these claims are explicitly written as “as claimed in claim #.”

3. Each claim mentions one or more entities (e.g., components, processes) that are preceded by determinants (‘a’, ‘an’, ‘the’).
4. Claims often mention the sub-entities of a mentioned entity through explicit hierarchical relationships like ‘comprising’, ‘having’, ‘including’ etc.
5. Claims do not include anaphoric references (e.g., ‘it’) to the entities that were mentioned in previous claims. Rather, entities are explicitly mentioned whenever needed.

The above properties allow us to extract facts as <entity, relationship, entity> triples from the claims of 6,004,075 utility patents as of 20th August 2019. We discard the design and plant patents in the work for our focus on engineering knowledge. The rule-based approach that we propose in this paper is repeatable, scalable, and adaptable within the US patent database.

## 3. Engineering Knowledge Graph

According to the overview shown in Figure 2, we extract facts in the form of <entity, relationship, entity> triples from each claim. We repeat the procedure for all claims and aggregate the facts in a patent document. We subsequently repeat the process for all patents in the source database. We explain the steps that are numbered within the arrows of Figure 2 as follows.

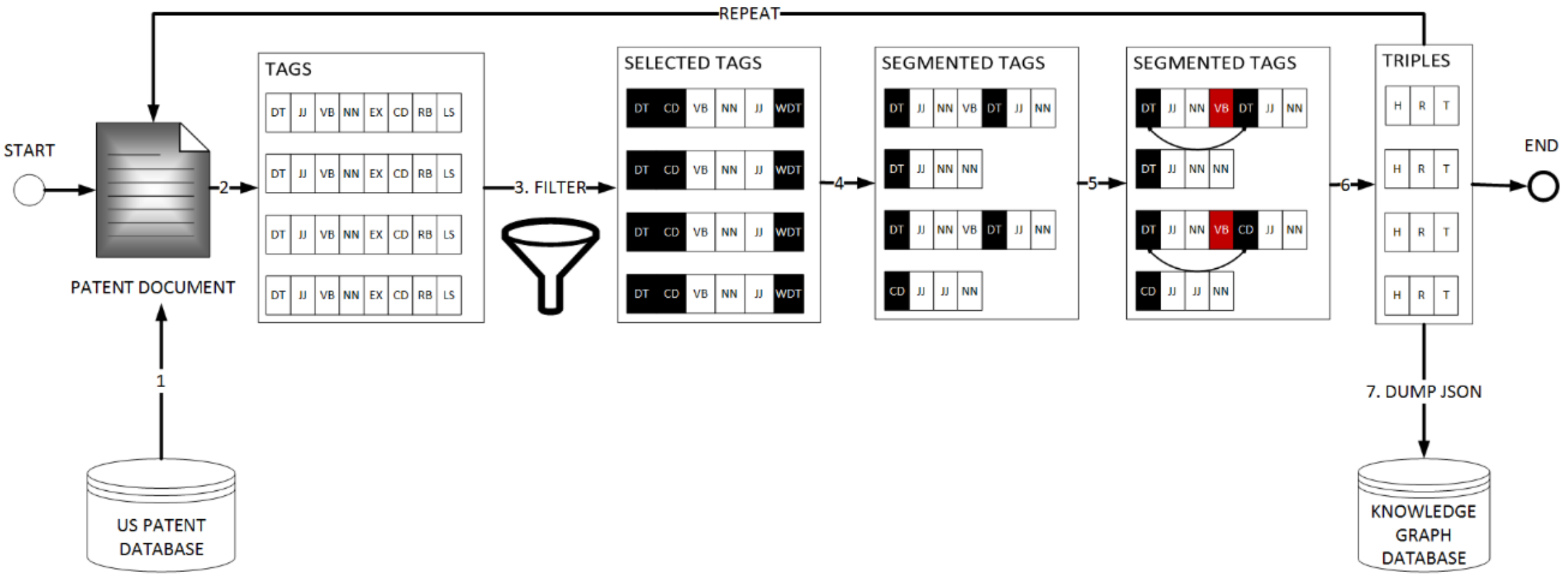

Figure 2: Overview of the proposed rule-based approach.

- Step 1 – First, we procure the claims in all patent document from the US patent database (Section 3.1).
- Steps 2, 3 – Second, we tokenise each claim along with Parts of Speech (POS) tags. Among the tags, we filter determinants, verbs, nouns, and adjectives (Section 3.2).
- Step 4, 5, 6 – Next, we segment the filtered tags at the positions where ‘which’ or ‘that’ are found. For each segment, we extract facts using the determinant markers and verbs that occur between these (Section 3.3).
- Step 7 – Finally, we aggregate the entities and facts for a patent document into the JavaScript Object Notation (JSON) format. We repeat this procedure for all patent documents (Section 3.4).

## 3.1. Collecting Data

We extract patent application numbers for all granted US patents from 1st January 1975 until 20th August 2019 from the source – PatentsView[11]. Then, we segregate these application numbers according to the year of application. For each batch comprising 10,000 patents, we crawl the GooglePatents[12] webpage using a web-crawler – BeautifulSoup[13].

[11] https://www.patentsview.org/download/

We append a given application number – 'US3230900A' to the base URL – "https://patents.google.com/" to reach the source webpage where we use the HTML class – 'claims' to retrieve the claims as an array of strings. We append these claims to a TSV file along with the application number. The data collection consumed over 300 hours, despite pooling 7-10 batches using a multi-processer. An alternative, efficient way is to directly download the claims (39 GB) from PatentsView, which we did not adopt due to resource limitations.

## 3.2. Cleaning Text

The purpose of this step is to transform a claim into a set of tokens, devoid of those that are not necessary for our work. Herein, we adopted the usual text cleaning procedure, which involves removal of punctuations, stop words, numbers, conversion to lower case etc. In addition to these, we repeat the following steps for each claim. It is important to note that each claim is made of a single sentence.

1. We remove the following words/phrases from the sentence. The following could be considered the data-specific stop words or phrases.
    a. The string '\n' occurs frequently due to HTML source code.
    b. The phrase 'as claimed in claim' denotes the dependency between claims, which is not important for our work.
    c. While frequent terms like 'claim, said, wherein, further' do not add value to our work, these are wrongly tagged as nouns that create issues in further steps.

[12] https://patents.google.com/
[13] https://www.crummy.com/software/BeautifulSoup/bs4/doc/

2. We identify Parts of Speech – POS tags using the Natural Language Toolkit – NLTK and preserve the order of tags as given in the source sentence.
    a. We filter tokens with the following tags whose definitions are given in Table 1: {NN, NNS, NNP, NNPS, DT, CD, JJ, JJR, JJS, VB, VBD, VBG, VBN, VBP, VBZ, and WDT}.
    b. The tags N**, J**, VB* denote nouns, adjectives, and verbs, respectively. While nouns and adjectives are tied to entities, verbs represent relationships [41], [42], [47].
    c. We also filter DT (a, an, the), CD (e.g., one, two), and WDT (which, that) that serve a significant role in further steps.
3. In the list of VB* tags, we identify the tokens that denote hierarchical relationship.
    a. These tokens were manually listed as [‘comprising’, ‘comprises’, ‘comprise’, ‘comprised’, ‘include’, ‘including’, ‘includes’, ‘included’, ‘consist’, ‘consists’, ‘consisted’, ‘consisting’, ‘has’, ‘having’]. Up to our knowledge, these words constitute an exhaustive set of verbs that denote system-subsystem relationship.
    b. We found that the NLTK POS tagger identified ‘comprise’ as a noun due to the neighbourhood tags. We override the tagger and set these words as verbs so that hierarchical relationships are tags accurately.

Table 1: Parts of Speech

| **Part of Speech** | |
|---|---|
| **CD** | cardinal digit |
| **DT** | determiner |
| **JJ** | adjective ‘big’ |
| **JJR** | adjective, comparative ‘bigger’ |
| **JJS** | adjective, superlative ‘biggest’ |
| **NN** | noun, singular ‘desk’ |
| **NNP** | proper noun, singular ‘Harrison’ |
| **NNPS** | proper noun, plural 'Americans' |
| **NNS** | noun plural 'desks' |
| **VB** | verb, base form take |
| **VBD** | verb, past tense took |
| **VBG** | verb, gerund/present participle taking |
| **VBN** | verb, past participle taken |
| **VBP** | verb, sing. present, non-3d take |
| **VBZ** | verb, 3rd person sing. present takes |
| **WDT** | wh-determiner which |

## 3.3. Extracting Facts

Entities in patent claims are always preceded by a determinant DT – 'a', 'an', and 'the' (See Figure 3a for an example). The claims also include, occasionally, the terms like 'that' or 'which' to refer to the immediately preceding entity ('DT' tag). For example, in Figure 3A, the second instance of 'that' refers to 'a peak wavelength'. Such instances of 'which' and 'that' are indicated as 'which determinant' – WDT by the POS tagger. We segment the claims at these instances to avoid anaphoric ambiguity as well as loss of entity recognition. To segment the claims, we mark the entity before WDT as the end of the previous segment and begin the next segment with the preceding entity, while removing the WDT token. We illustrate this step using the example shown in Figures 3A and 3B.

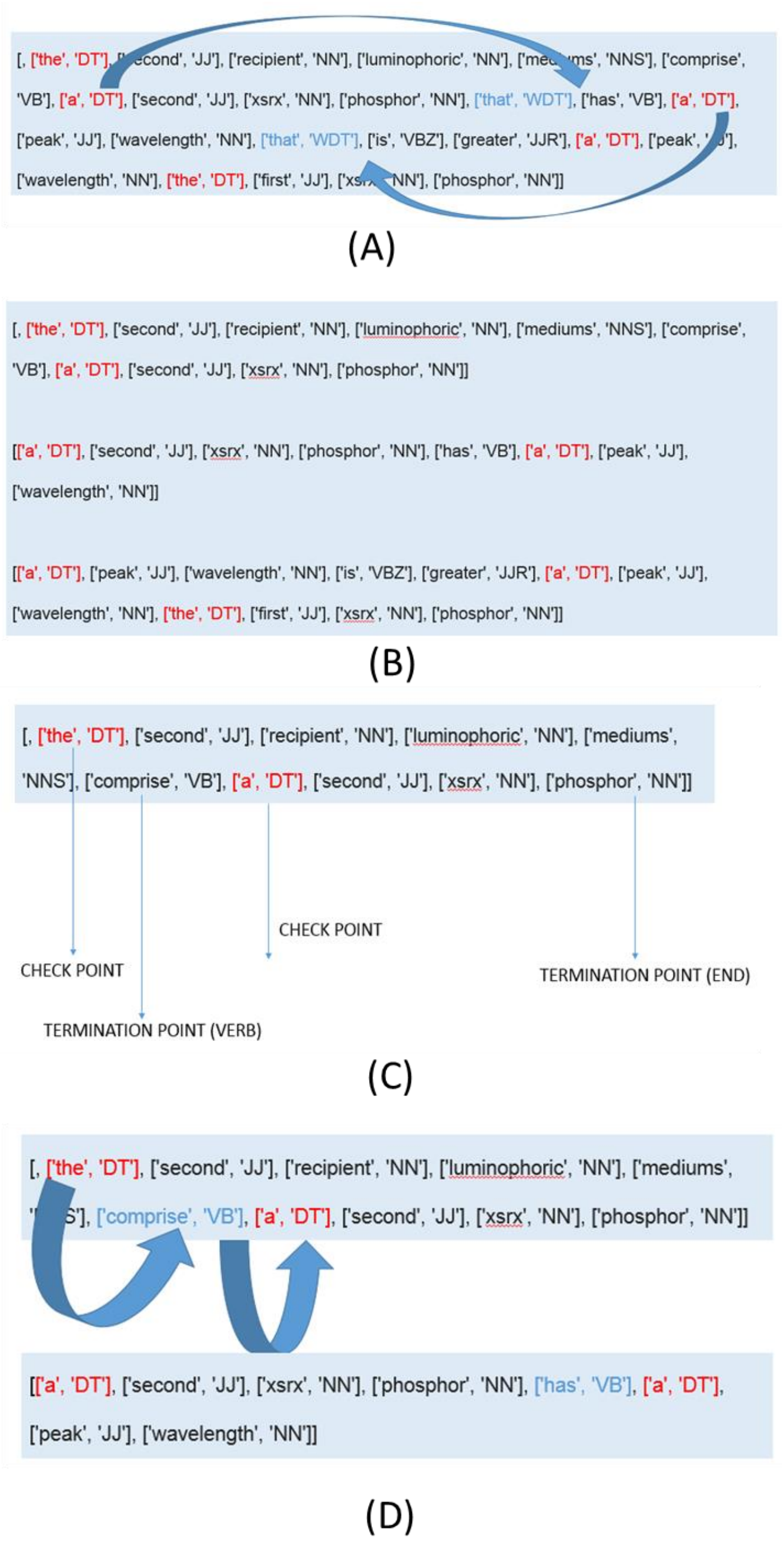

Figure 3: Tags A) before and B) after segmentation; C) termination points for entities; D) extracting relationships.

In each segment, we extract the entities as a set of nouns and adjectives that follow a <DT, CD> determinant. We then combine the set of words into a single string in the exact order that was found in the source. While cleaning the claims (Section 3.2), we preserve the order of tags to serve this need. Combining words into a single term serve a meaningful purpose as acknowledged in the literature [48], [49].

Multi-word terms are more specific and less ambiguous than single-word terms [49]. In the patent analysis, one could split a sentence into a series of words and then use merge, drop, and accept rules to get back the combined version of words. As a result, the term “public high school” could be retrieved *verbatim*, so that the term serves the meaning that the context has to offer, as well as bereft of the ambiguities that arise from individual words such as ‘public’, ‘high’, and ‘school’ [48].

Relationships could be extracted as verbs, which often exhibit polysemy or mutability; i.e., they refer to multiple meanings in different contexts [49]. Hence, capturing verbs along with the neighbourhood entities provide us with adequate context to infer the meaning. The verbs extracted from patents have been classified into functional, structural, logical, and non-technical [50]. Since claims do not provide information apart from structure, the relationships are largely structure-oriented, e.g., comprising, connected, adapted etc. There are few instances where the relationships could denote the behaviour of the system, e.g., move, push etc.

Once the DT, CD markers are identified in a segment, we loop through tags until a termination point, which could be the following tags: DT, CD, VB, or the end of the tag list (See Figure 3C). We extract the subset of the tag list and concatenate these into a single string. For instance, the entities extracted for the running example are as follows: [‘the second recipient luminophoric mediums’, ‘a second xsrx phosphor’, ‘a peak wavelength’, and ‘the first xsrx phosphor’].

From the extracted strings, we remove {‘the’, ‘a’, ‘an’} – DT, but we do not remove count – CD as these accommodate useful information [48]. For example, ‘second xsrx phosphor’ is different from ‘first xsrx phosphor’. While we do not set an upper bound on the length of an entity, the number of entities is restricted to the number of DT and CD tags. Using this upper bound, we would be able to check if all the entities are extracted.

As shown in Figure 3D, the relationships are extracted as VB tags between any two entities. These are classified as hierarchical and non-hierarchical, depending on the list of verbs mentioned in Section 3.2. We capture multiple sub-entities if the claim has the following format: ‘A comprises B, C, and D’. We also capture multiple levels of the hierarchy if the claim is expressed as ‘A comprises B and C, which further comprises D, E, and F’. Such levels are captured through segmenting claims as follows: ‘A comprises B and C’, ‘C comprises D, E, and F’. We repeat the procedure for all claims present in all patents in the TSV file (output of the extracting claims in Section 3.1).

## 3.4. Structuring Facts

We organise the facts obtained from each claim into a dictionary as shown in Figure 4A, where the primary key is the application number. The secondary keys include entities and relationships. Each relationship is a fact or triple that includes source entity, relationship, target entity. We use JSON dumps () to store the whole dictionary into a single text file that can be parsed later if required.

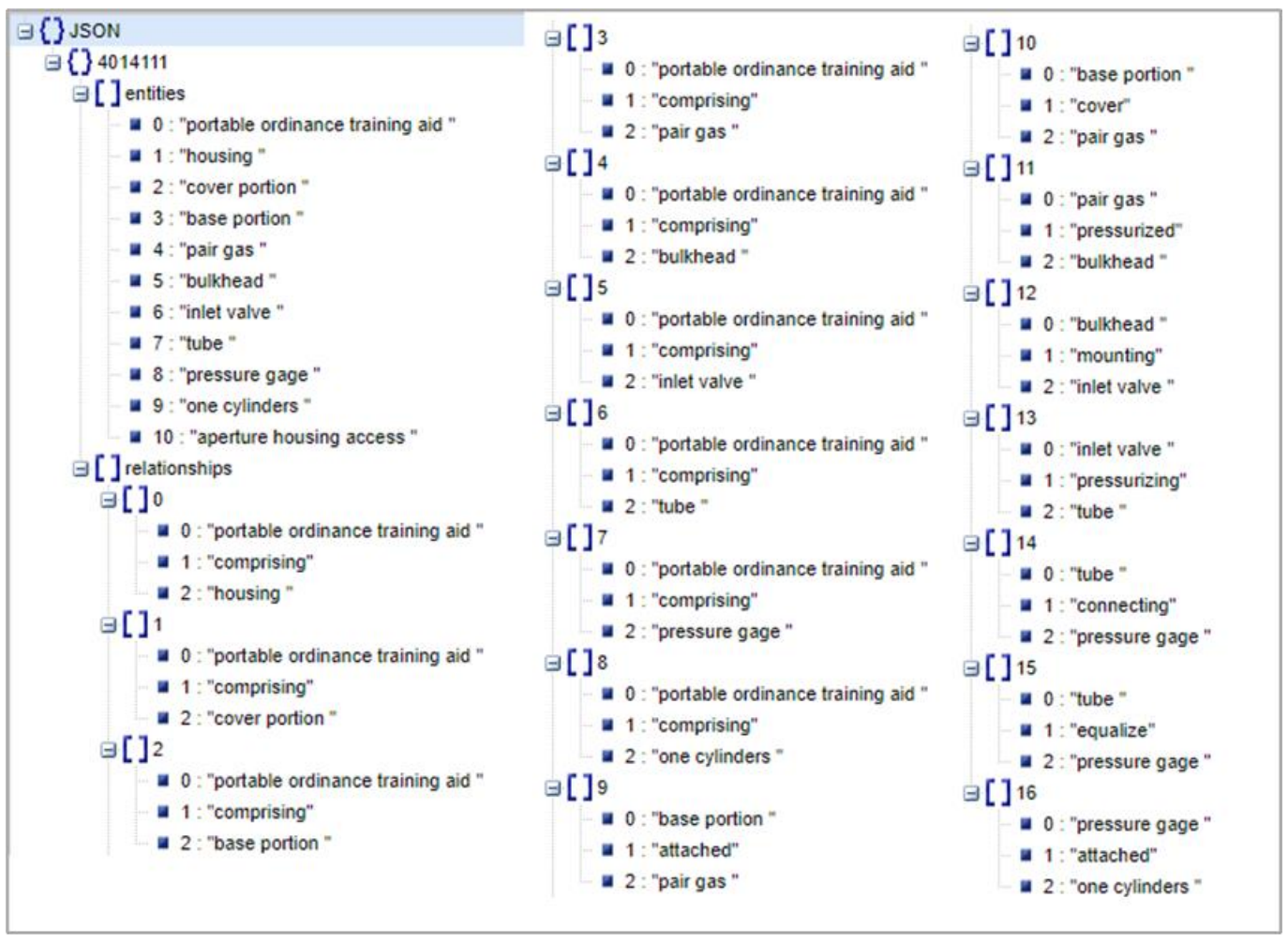

(A)

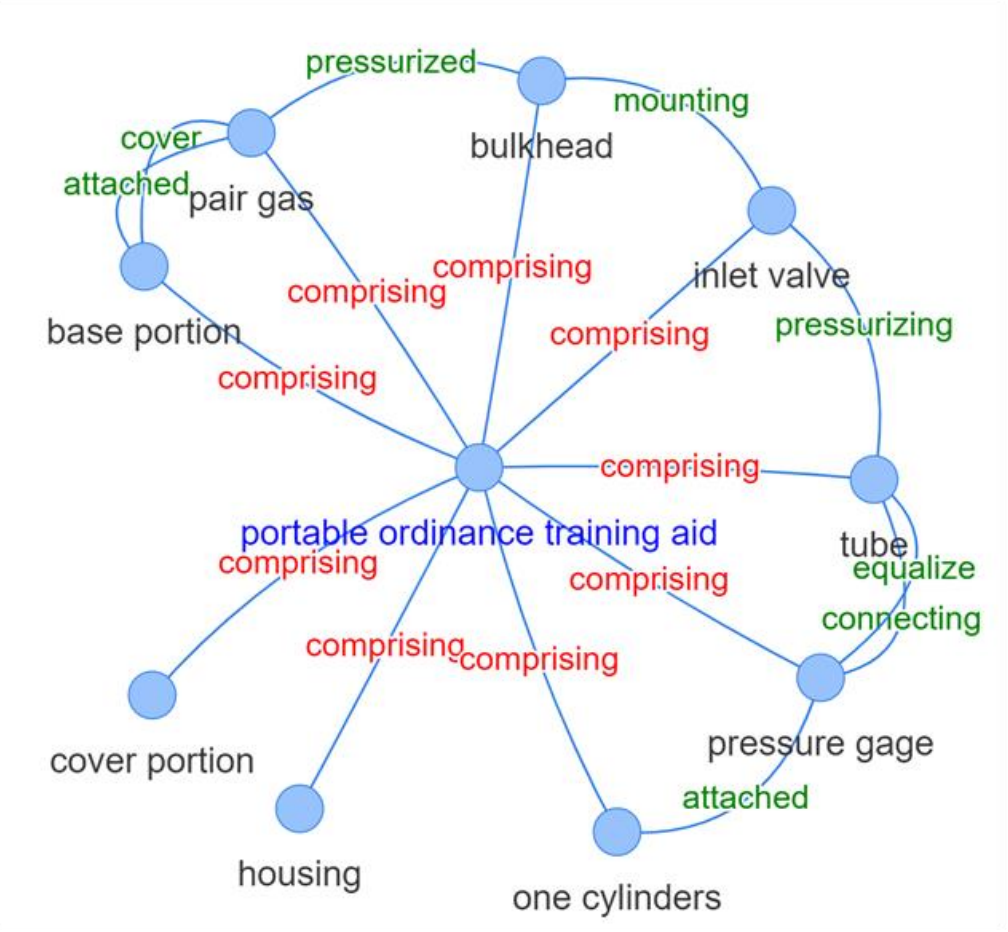

(B)

Figure 4: The knowledge graph of US4014111 – "Ordnance Training Aid" A) stored in JSON format, B) visualised using a graph, where Central Topic - ■, Entity - ■, Hierarchical Relationship - ■, Non-hierarchical Relationship - ■

We segregate the text files including dictionaries for a batch of 10,000 patents according to the application year (1975-2018). The database constitutes 706 text files consuming over 49 GB. For illustration, we represent a patent knowledge graph in both dictionary (Figure 4A) and graph (Figure 4B) forms. Overall, according to Figure 5, we have converted the patent database from pure text format to a structured graph of interconnected facts that are represented as triples.

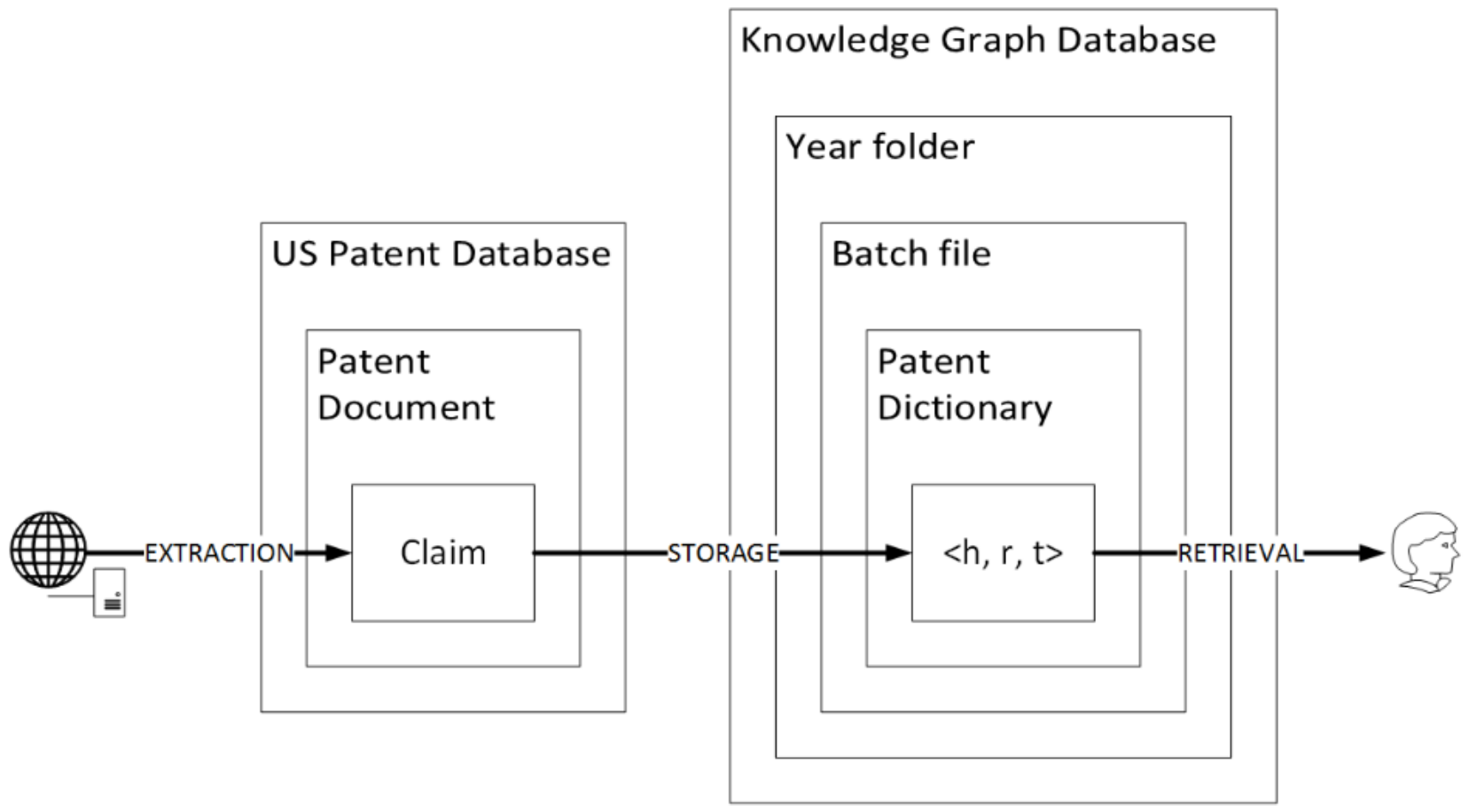

Figure 5: The schemata of our knowledge graph database.

## 3.5. Reasoning Illustration

In this section, we illustrate reasoning over the engineering knowledge graph using an example of an oximeter, which is a recently emerged personal health care device that was designed to measure the level of oxygen in the least evasive manner. We extract the neighbourhood knowledge graph surrounding the entity – ‘oximeter’ and retrieved 8,442 facts. Among these, we arbitrarily selected a few facts and display the graph as shown in Figure 6. In Figures 6A and 6B, we display all entities but restrict these to, respectively, hierarchical, and non-hierarchical relationships. In a pure hierarchy knowledge graph like Figure 6A, an applicable existential rule is shown in the first-order representation as follows.

$$\forall x \forall y \forall z \; comprises(x,y) \wedge comprises\,(y,z) \;\; \rightarrow \;\; comprises(x,z) \tag{1}$$

The rules like above form the pillar of reasoning over the knowledge graph. The scope of an existential rule could vary from a specific product domain to engineering as a whole. In the context of input-output transformations, Mao and Sen [51], [52] propose a reasoning algorithm that infers flow types, suitable functions and topologies. The algorithm is supported by an ontology that includes the following: 1) abstract classes for common material and energy flows (e.g., solid, liquid), 2) common flow attributes (e.g., hot, cold) that are categorised into pressure, temperature, and volume, 3) a map between functions (energy, work) and flow attributes using a correlation matrix that was built using qualitative physics.

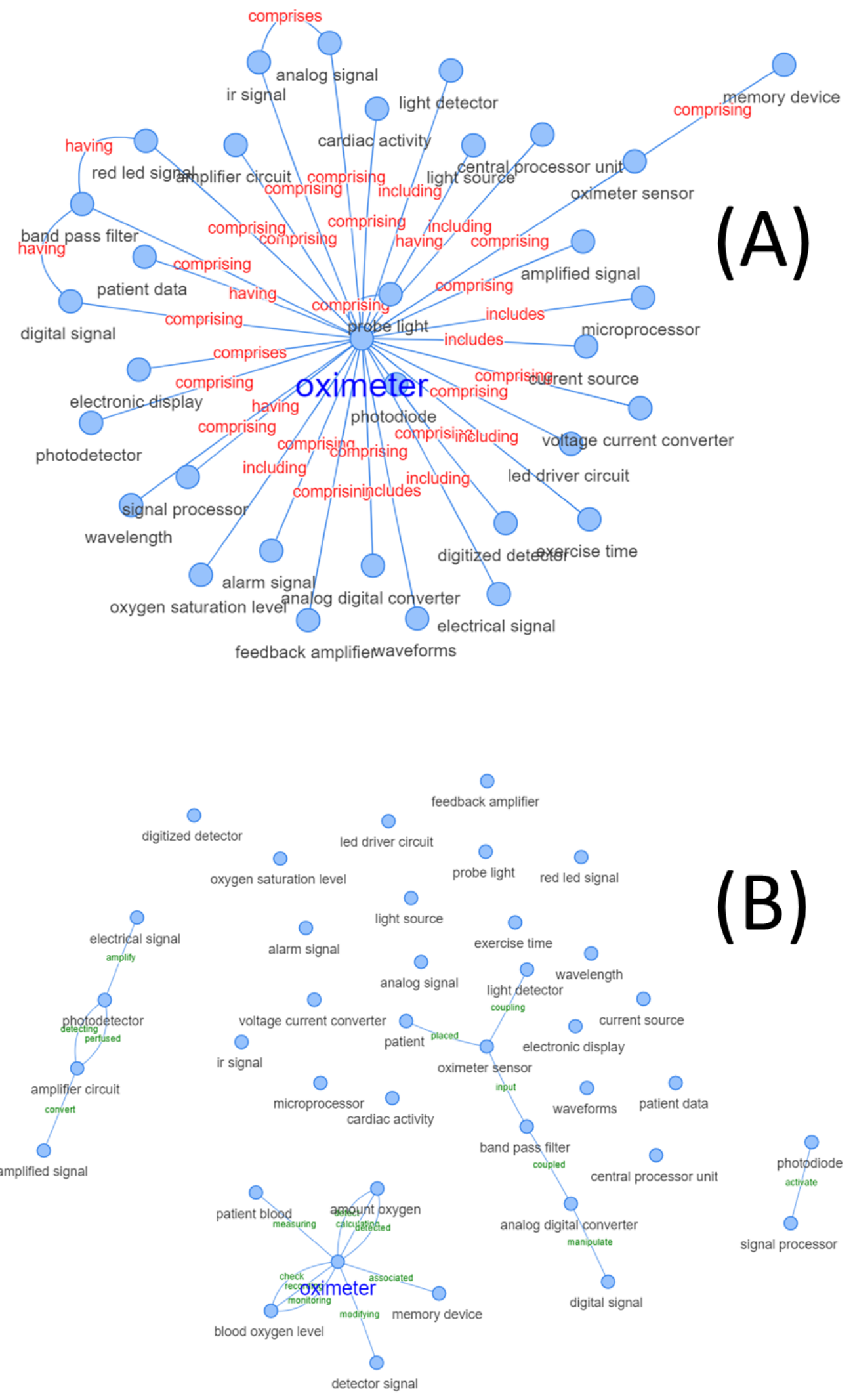

Figure 6: The neighbourhood knowledge graph of the oximeter. A) hierarchical and B) non-hierarchical relationships

Such generalised reasoning approaches coupled with domain-specific existential rules could enable rich possibilities of reasoning over the knowledge graph that we have extracted. For example, in Figure 6B, since it is understood that ‘analog digital converter’ transforms analog signal to digital signal, we could infer two additional relationships as follows: <analog digital converter, manipulate, analog signal>, <analog signal, convert, digital signal>. Since ‘analog signal’ comprises ‘ir signal’ as shown in Figure 6A, we could also infer the following: <analog digital converter, manipulate, ir signal>, <ir signal, convert, digital signal>.

For such inferences to be enabled, a supporting ontology (e.g., Mao and Sen [51]) must be built. Such an ontology should include classes (e.g., entity classification in Google Knowledge Graph[14]) and attributes that characterize each entity and relationship, e.g., attributes of an analog-digital converter. The ontology plays a crucial role in building Q/A, query, and recommendation systems (e.g., Microsoft Academic Graph) that require named entity recognition and relation linking.

For example, a general term like ‘electrical signal’ could be present in various contexts in our knowledge graph. It is important to recognize the specific context by processing query using ontologies as supports. For such recognition, entities and relationships should be assigned a unique ID and linked to their classes and attributes. In this context, an entity could be linked to a patent number, which could be further linked to the source patent data for retrieving domain name, year of application, inventors etc. Such linking would allow us to focus the search on specific parts of the knowledge graph and answer specific questions like “which components were introduced in an oximeter after 2018?”.

[14] https://developers.google.com/knowledge-graph

# 4. Evaluation

We report a comparison (Section 4.1) of a small portion (30 facts) of our knowledge graph against a similar portion of triples obtained from TechNet[15] and ConceptNet, both of which are publicly accessible via their APIs. We also report the size and coverage of our knowledge graph against some well-known benchmarks (Section 4.2).

## 4.1. Qualitative Comparison

ConceptNet is a common-sense knowledge graph comprising facts obtained using DBpedia and Open Mind Common Sense projects [23]. TechNet is a technological concept network that provides relevance scores among 4 million entities obtained from patent titles and abstracts [53]. While ConceptNet is a comparable benchmark for the nature of relationships in our work, TechNet includes entities from the patent database, which is also our source of knowledge.

A central problem of evaluating a knowledge base is to align the entities along with the ground truth [54]. In this evaluation, we set a central topic – 'hairdryer' and obtained triples surrounding these as follows.

1. ConceptNet – We obtain 30 surrounding entities using the ConceptNet API[16] – (See Figure 7A) and second level entities extending from five of them (Figure 7B).
2. TechNet – We use the TechNet API[17] to obtain 30 entities that have the strongest relevance to 'hairdryer' (Figure 8A) and second level entities extending from five of them (Figure 8B).
3. Our engineering knowledge graph – We searched for facts $< h, r, t >$ where the $h$-head entity exactly matches 'hairdryer'. We arbitrarily selected 30 relationships from over

[15] http://www.tech-net.org/
[16] http://api.conceptnet.io/c/en/
[17] https://github.com/SerhadS/TechNet/tree/master/APIs

16,000 relationships (See Figure 9A) and second level entities by extending five of them (Figure 9B).

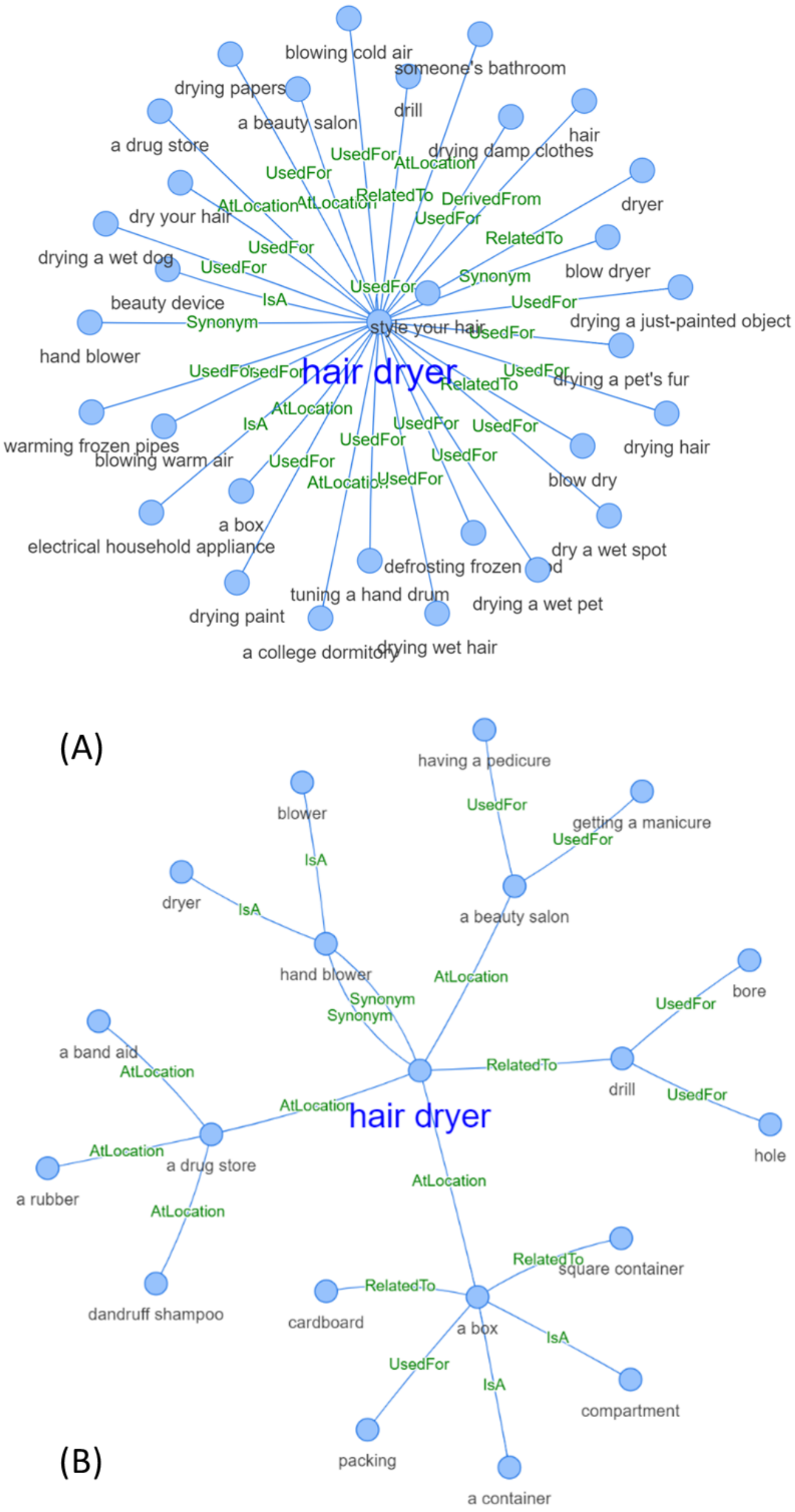

Figure 7: A) First Level and B) Second Level entities and relationships around the central entity “hairdryer” in ConceptNet. Legend: Central Topic - ■, Entity - ■, Qualitative Relationship - ■Legend: Central Topic - ■, Concept - ■, Qualitative Relationship - ■

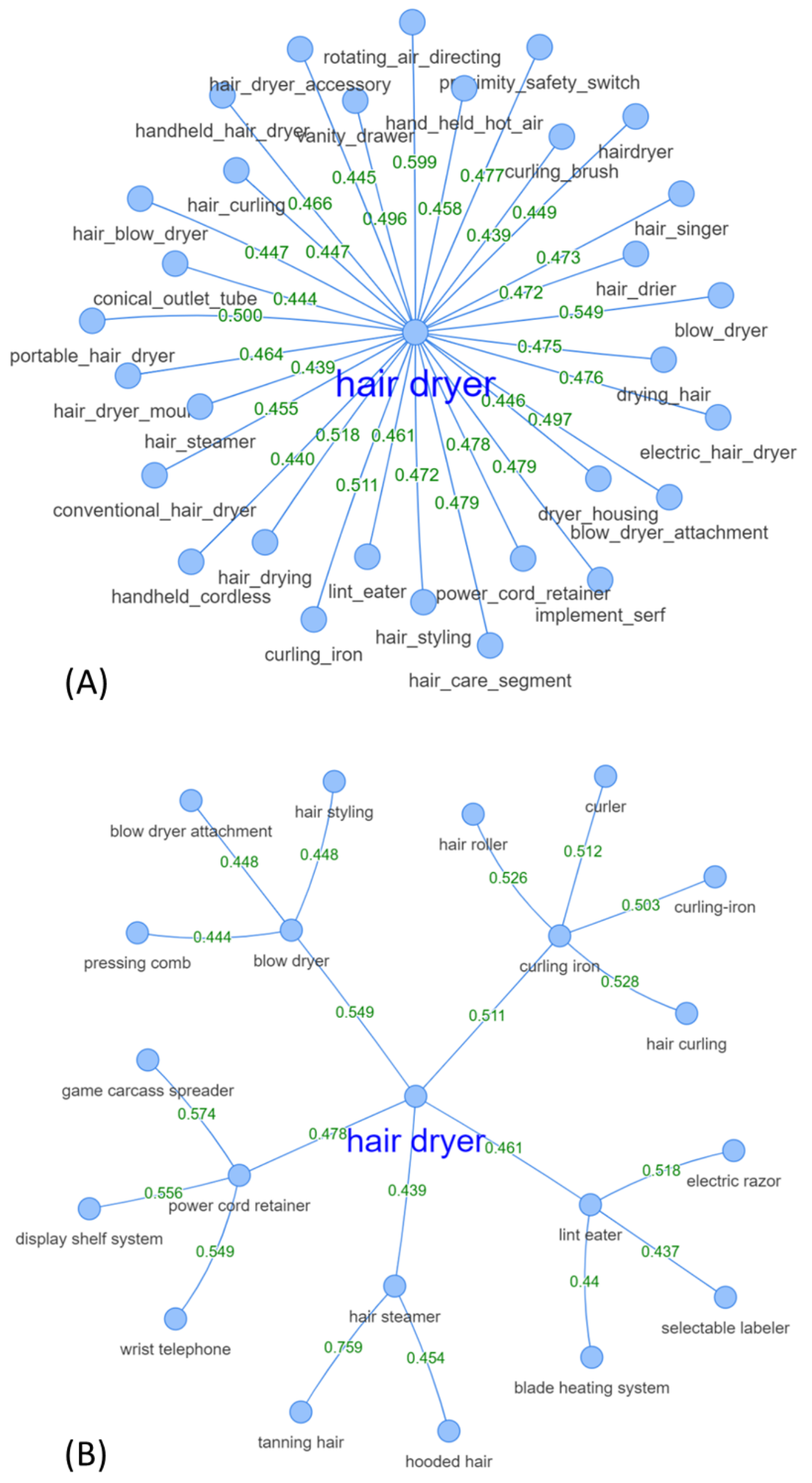

Figure 8: A) First Level and B) Second Level entities and relationships around the central entity “hairdryer” in TechNet. Legend: Central Topic - ■, Entity - ■, Qualitative Relationship - ■Legend: Central Topic - ■, Concept - ■, Qualitative Relationship - ■

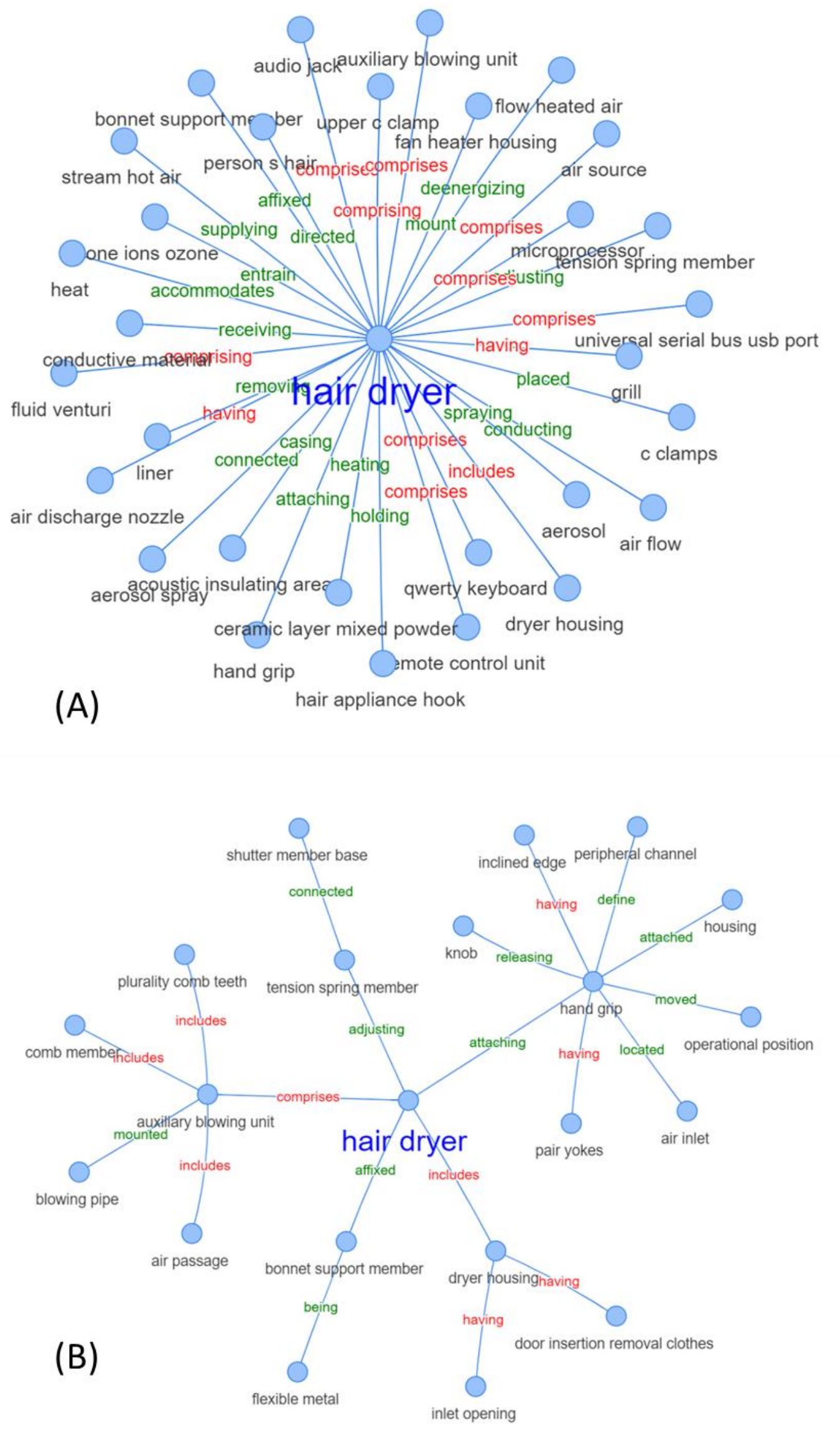

Figure 9: A) First Level and B) Second Level entities and relationships around the central entity “hairdryer” in our knowledge graph. Legend: Central Topic - ■, Entity - ■, Qualitative Relationship - ■Legend: Central Topic - ■, Concept - ■, Qualitative Relationship - ■

As shown in Figure 7A, some of the relationships in ConceptNet include the following: *related to, is a, part of, member of, has a, used for, capable of, at location, causes,* and *synonym*. It is important to note that ConceptNet was created based on the mining of only 21 types of predefined common-sense relations [23], whereas the relationships in our knowledge graph (example in Figure 9A) do not fall into specific categories and appear more diverse. Apart from hierarchical relationships (■), there are several relationship labels like ‘receiving’, ‘connected’, ‘energized’, ‘mating’, ‘attaching’, etc. The relationships in TechNet, as shown in Figure 8A, are quantitative and the network is not qualified as a knowledge graph.

Possibly due to limited pre-processing, unlike our work, ConceptNet seems to treat entities such as ‘hairdryer’ and ‘a hairdryer’ differently [55]. The overall difference between Figure 7A and 9A is that ConceptNet includes common-sense ontological relationships [56], and our work includes domain-specific technical relationships that are extracted directly from the text. The relationships in ConceptNet, except *part of* add little or no value to application in engineering design. For instance, <hairdryer, *used for,* dry your hair> and <hairdryer, *at location*, a beauty salon> provide mere obvious information. On the other hand, in Figure 9A, relationships like <hair dryer, *spraying*, aerosol> and <hair dryer, *includes*, drying housing> provide purely technical information.

ConceptNet shows a concept as a combination of verbs, articles, adjectives, and nouns [57], e.g., ‘drying a wet dog’ that is better understood as an action rather than an elemental concept. In a knowledge graph, the phrase is better formulated as <xxx, drying, wet dog> where ‘xxx’ and ‘wet dog’ could be treated as a pair of concepts that are contextualised by a relation – ‘drying’. ConceptNet does not take into account such rich information as we have, in our rule-based extraction of knowledge graphs. To de-contextualize a concept, it must be composed of just noun-noun (e.g., dog castle) or adjective-noun (e.g., wet dog) combinations [58], given word sense disambiguation is taken into account.

We extract second-level entities for ‘hand grip’ and ‘auxiliary blowing unit’ in Figure 9B. Upon doing so, we can merge facts from multiple patents into a single integrated knowledge graph. Therefore, if a high-level entity is given, we could obtain its constituents at multiple levels from multiple patents, e.g., <hairdryer, includes, auxiliary blowing unit>, <auxiliary blowing unit, includes, comb member>, <auxiliary blowing unit, includes, air passage>.

Although the TechNet (Figure 8) is patent-based, entities like 'hair_singer', 'hair_curling' denote the purpose, which is usually mentioned in the titles and abstracts. Patent claims are free from these non-technical terms, as a result, these were not found anywhere in our work. While a quantitative relationship serves several statistical benefits, it may not, *per se*, tell us how entities could be connected.

While ConceptNet provides common-sense facts, TechNet provides only quantitative relationships among entities. These knowledge bases have been reported to support ideation, as these are utilised to expand entities surrounding a target entity. In addition, they have been demonstrated to support design knowledge representation, as they could be used to visualize the structural relationships between entities in a design [10], [32], [59]. In contrast, our knowledge graph provides engineering entity-relationship-entity facts, which could further augment engineering knowledge representation, reasoning, and retrieval. These knowledge bases should therefore complement each other.

## 4.2. Quantitative Comparison

To assess the coverage of terms (i.e., entities) in our knowledge graph, we utilise the MIT-Cambridge Engineering Dictionary[18] that includes 2,704 representative terms across six fields of engineering: civil and structural, material, mechanical, mining, nuclear, and software. The coverage in each of these fields is compared against some well-known databases as shown in Table 2. Among the entries in Table 2, WordNet is a lexical database for the English language and ConceptNet is a common-sense knowledge base. Word2vec [40] and GloVe [60] are word embedding models that were trained on Google News, Wikipedia, gigaword, and Common Crawl as the original data sources. The entries in Table 2 that mention these models are the resultant semantic networks. B-Link [61] and TechNet [53] are semantic networks that were developed from, respectively, ScienceDirect[19] and patent database. The comparison shows that our knowledge graph has the best term coverage of 80.7% across fields.

[18] http://www-mdp.eng.cam.ac.uk/web/library/enginfo/mdpdatabooks/dictionary1.pdf

Table 2: The proportion of engineering dictionary terms found in different databases. The last two rows correspond to our work.

| | **Civil** | **Material** | **Mech** | **Mining** | **Nuclear** | **Software** | **Total** |
|---|---|---|---|---|---|---|---|
| WordNet | 0.494 | 0.534 | 0.440 | 0.557 | 0.374 | 0.325 | 0.398 |
| ConceptNet | 0.685 | 0.686 | 0.632 | 0.668 | 0.591 | 0.629 | 0.637 |
| B-Link | 0.764 | 0.723 | 0.627 | 0.546 | 0.527 | 0.504 | 0.553 |
| Word2Vec | 0.449 | 0.496 | 0.446 | 0.590 | 0.372 | 0.463 | 0.470 |
| GloVe | 0.449 | 0.458 | 0.440 | 0.563 | 0.342 | 0.576 | 0.567 |
| TechNet | 0.876 | 0.841 | 0.761 | 0.799 | 0.698 | 0.671 | 0.723 |
| This work | 0.899 | 0.898 | 0.833 | 0.870 | 0.783 | 0.77 | 0.807 |
| This work (Adjusted) | 0.955 | 0.913 | 0.861 | 0.883 | 0.816 | 0.786 | 0.827 |

Exactly 522 terms were not found in our knowledge graph. Upon re-examining, we understood that these terms were absent due to many cleaning steps (Section 3.2) we performed on raw text data. For instance, ‘center of gravity’ was absent, as we exclude prepositions from original data. Similarly, while stripping punctuations, ‘x.225’ was changed to ‘x 225’. Additional examples are shown in Table 3. After processing these terms, we performed the search again.

[19] https://dev.elsevier.com/

Table 4: Variety of cases identified when a term was not found in our graph database.

| Reason | Original Term | Transformed Term |
|---|---|---|
| Use of prepositions | center of gravity | center gravity |
| Use of punctuations | x.225 | x |
| Use of operators | tcp/ip | tcp ip |
| Use of upper case characters | Clean coal | clean |
| Use of numbers | dod std 2168 | dod std |
| Joint words depicting a name | ObjectWorks | object works |

We observe that not only the exemplar terms but also additional 50 terms were found. It is not that the additional 50 terms were absent in our database; instead, these were present in a different form. Therefore, the updated term coverage is 82.7% (Table 3), which is significantly better than the previous best – TechNet. It is important to note that TechNet was built using mere Titles-Abstracts and ours from claims. The purpose of title, abstract, claims and descriptions are different and these contain different keywords [15]; possibly this is why our term coverage is higher.

Besides, in direct comparison with other databases (Table 5), our knowledge graph has 288 million entities and 795 million relationships among these. Each entity constitutes approximately 2.75 relationships. Lu et al. [62] specify that databases with more than one million entities are qualified as 'big knowledge' sources. The connectivity should be such that the number of edges is greater than one billion. By that specification, from Table 5, none of the databases is strictly qualified as a 'big' knowledge base. Our work and ConceptNet satisfy the criterion for entities and relationships, respectively.

Table 5: Comparing the size of different databases

| Databases | Number of Entities | Number of Relationships |
|---|---|---|
| TechNet | 4,038,924 | $\sim 8.5 \times 10^{12}$ |
| WordNet | 155,236 | 647,964 |
| ConceptNet | 516,782 | $\sim 1.3 \times 10^{11}$ |
| Word2Vec | 3,000,000 | $\sim 4.5 \times 10^{12}$ |
| GloVe | 417,194 | $\sim 1.7 \times 10^{11}$ |
| B-Link | 536,507 | 3,726,904 |
| This work | 288,807,731 | 794,956,771 |

## 5. Discussion

Our work addresses the growing demands for large scalable knowledge graphs to support engineering knowledge retrieval, representation, and concept generation-cum-evaluation, and the limitations posed by the reliance on common-sense knowledge graphs. We exploit the syntactic properties of patent claims to extract engineering facts <h, r, t> that are integrated and organised into a large engineering knowledge graph suitable for engineering design retrieval, representation, reasoning, and inference. The extraction method purely relies on a set of rules that are repeatable, scalable, and adaptable within the US Patent Database. It provides facts that include technical, qualitative relationships in contrast to ConceptNet and TechNet. It also has a greater size and coverage compared to the previous knowledge bases.

Our knowledge graph could potentially lead to several research opportunities. First, several embedding and classification algorithms could be implemented for node classification and link completion [20]. Second, the knowledge graph coupled with reasoning methods paves way for the development of Q/A and recommendation systems for engineering applications. Third, engineering scholars could utilise the knowledge graph as supports in the design process and examine its efficacy as a reasoning and knowledge aid.

Upon the development of efficient search methods on the engineering knowledge graph, we could couple these with problem-solving approaches in engineering design such as design-by-analogy [63], mind-mapping [10], TRIZ [64]–[66] etc. In TRIZ, for example, if we choose a pair of contradiction such that we wish to improve *reliability* by preserving *speed*, the guide recommends a change of temperature. We could filter facts from our knowledge graph that include ‘temperature’ in the entities and further converge by choosing the domains, entity type etc. The efficacy of such approaches to support the design process in various conditions could be studied in the future.

Our work, as reported in this article, is merely the initial progress towards building the engineering knowledge graph and the entailing applications for knowledge-based AI in engineering. As we move forward, we could explore new NLP techniques and other engineering data sources (e.g., engineering research articles). We hope the readers view this paper as an invitation for more research on the construction and applications of the engineering knowledge graph.